\documentclass[twocolumn]{jpsj2} 
\newcommand{\G}{\Gamma}
\newcommand{\dg}{\dagger}
\newcommand{\up}{\uparrow}
\newcommand{\dw}{\downarrow}
\title{Realization of Strong Coupling Fixed Point in Multilevel Kondo Models}

\author{Kazumasa \textsc{HATTORI}, Yosuke \textsc{HIRAYAMA} and Kazumasa \textsc{MIYAKE}}

\inst{Division of Materials Physics, Department of Materials Engineering Science, Graduate School of Engineering Science, Osaka University, Toyonaka, Osaka 560-8531, Japan}

\abst{Impurity four- and six-level Kondo model, in which an ion is tunneling among four- and six-stable points and interacting with surrounding conduction electrons, are investigated by using the perturbative and numerical renormalization group methods. It is shown that purely orbital Kondo effects occur at low temperatures in these systems which are direct generalizations of the Kondo effect in the so-called two-level system. This result offers a good explanation for the enhanced and magnetically robust Sommerfeld coefficient observed in SmOs$_4$Sb$_{12}$ and some other filled-skutterudites.}

\kword{Kondo effect, rattling, multi-level system}

\begin{document}
\maketitle
Since the two-level Kondo effect was studied about three decades ago,\cite{Kondo1,Kondo2} the physics arising from the off-center degrees of freedom of ions have attracted much attention.
 Recently, off-center motions of ions in clathrate compounds\cite{NemotoCePd} and  filled skutterudites\cite{GotoPrOs} have been identified by ultrasonics. One of the recent remarkable experimental results is a realization of magnetically robust heavy fermion in SmOs$_4$Sb$_{12}$, in which the linear temperature coefficient of the specific heat at low temperatures amounts to 820mJ/(mol$\cdot$K$^2$) even up to 8-15 T.\cite{Sanada,Yuhasz} Since the Sm ion is considered to be in a mixed valence state, it is mysterious why such a heavy effective mass is realized. 

Characteristic properties of these materials are the existence of thermally excited rattling modes if an ion in the ``cage''. In the case of PrOs$_4$Sb$_{12}$, it is claimed that there exist degenerate $\Gamma_3^{+}$ (cubic $O_h$) modes of the off-center motion of the ion (Pr$^{3+}$) to explain the ultrasonic dispersion in $C_{11}-C_{12}$. (This point has been still controversial. There are two different results in $C_{44}$\cite{GotoPrOs, Nakanishi1}). However, textbooks for quantum mechanics tell us that a ground state is a singlet with a symmetric wavefunction and there exists no node in a spinless one-body problem. From this view point, the interpretation above is not easily accepted.

 In this paper, we investigate an impurity four- and six-level Kondo model, in which an ion is tunneling among four- and six-stable points. We take into account interactions between the ion and the surrounding conduction electrons, and use the perturbative renormalization group (PRG) method to see low energy properties of the system. Since we want to investigate the effects arising from the non-magnetic origin, we assume the conduction electron to be spinless. We restrict ourselves mainly to the six-level model (see ref. 8 for the detailed PRG and numerical renormalization group (NRG) discussions of four-level model). We will make a brief comment about the results of the four-level model in the last part.

In order to make the model simple, we consider a Gaussian wavefunction of the ion $\phi_i^0({\bf x})$ at each off-center position $i$, 
\begin{eqnarray}
 \phi_i^0({\bf x})= \Big(\frac{1}{\pi\sigma}\Big)^{\frac{3}{4}}\exp\Big[-\frac{({\bf x}-{\bf x}_i)^2}{2\sigma^2}\Big],
\end{eqnarray}
where $\sigma$ and ${\bf x}_i$ are the width of the Gaussian and coordinate of off-center positions ${\bf x}_i=(\pm a,0,0),\ (0,\pm a,0)\ {\rm and}\ (0,0,\pm a)$, respectively ($a$ and $\sigma$ are measured in the unit of inverse Fermi wavenumber $k_F^{-1}$ ). Using $\phi_i^0({\bf x})$, we can evaluate the transfer integral $S$ directly (we keep only nearest neighbor terms). Since $S=\exp(-a^2/(2\sigma^2))\not=0$, the energy eigenstates of the ion are classified by each point group. Then, the non-interacting Hamiltonian $H_0$ is given as
\begin{eqnarray}
H_0=\sum_{\mu}\Delta_{\mu}a^{\dag}_{\mu}a_{\mu}+\sum_{k}\sum_{\hat{l}}\epsilon_{k\hat{l}}c^{\dag}_{k\hat{l}}c_{k\hat{l}},
\end{eqnarray}
 where $a_{\mu}$ is the annihilation operator of the ion with the symmetry index $\mu$ and energy $\Delta_{\mu}$ and $c_{k\hat{l}}$ is annihilation operator of the conduction electrons with the radial wavenumber $k$, angular momentum $\hat{l}=(l,m)$, and energy $\epsilon_{k\hat{l}}$. For the ion operators, the constraint $\sum_{\mu}a_{\mu}^{\dagger}a_{\mu}=1$ is required.

The form of the interaction part of Hamiltonian $H_{\rm int}$ is expressed in general as{\cite{HHM}}
\begin{eqnarray}
H_{\rm int} &=& \sum_{kk'}\sum_{\hat{l}\hat{l'}}\sum_{\mu\nu}\sum_{\gamma} J_{\hat{l}\hat{l'}}^{\gamma}c_{k\hat{l}}^{\dag}c_{k'\hat{l'}}a_{\mu}^{\dag}{\bf \Psi}^{\gamma}_{\mu\nu}a_{\nu},
\end{eqnarray}
where $J_{\hat{l}\hat{l'}}^{\gamma}$ are coupling constants and ${\bf \Psi}^{\gamma}_{\mu\nu}$ are matrices of the $\gamma$-th irreducible representation (IR) of the direct product $a^{\dag}_{\mu}a_{\nu}$ and subject to the constraint as $\sum_{\mu\nu}|\Psi^{\gamma}_{\mu\nu}|^2=1$.

 For this kind of Hamiltonian $H=H_0+H_{\rm int}$, 2-loop renormalization group (RG) equations have been derived formally in ref. 9
.  A set of RG equations is given as follows:
\begin{eqnarray}
\!\!\!\!\!\!\!\!\!\!\!\frac{\partial J_{\hat{l}\hat{l'}}^{\gamma}}{\partial x}\!\!\!\!\!&=& \!\!\!\!\!\rho\sum_{\hat{l''}}\sum_{\alpha\beta}J_{\hat{l}\hat{l''}}^{\alpha}J_{\hat{l''}\hat{l'}}^{\beta}{\rm Tr}\Big([{\bf \Psi}^{\alpha}, {\bf \Psi}^{\beta}] {{\bf \Psi}^{\gamma}}^{\dag}\Big)\nonumber\\
\!\!\!\!\!&+&\!\!\!\!\!\sum_{\alpha\beta\lambda}\frac{\rho^2}{2}{\rm Tr}\Big(J^{\alpha}J^{\beta}\Big)J_{\hat{l}\hat{l'}}^{\lambda}{\rm Tr}\Big([{\bf \Psi}^{\alpha},[{\bf \Psi}^{\beta}, {\bf \Psi}^{\lambda}]]{{\bf \Psi}^{\gamma}}^{\dag}\Big),\label{RG1}\\
\!\!\!\!\!\!\!\!\!\!\!\frac{\partial \Delta_{\mu}}{\partial x}\!\!\!\!\!&=&
-\rho^2\sum_{\alpha\beta}{\rm Tr}\Big(J^{\alpha}J^{\beta}\Big)\sum_{\nu\not=\mu}{\bf \Psi}_{\mu\nu}^{\alpha}{\bf \Psi}_{\nu\mu}^{\beta}\Delta_{\nu}\label{RG2},
\end{eqnarray}
where $x=\log (D/D_0)$, $D$ $(D_0)$ being the half of the scaled (bare) conduction-electron bandwidth, and $\rho$ is the density of states (DOS) at the Fermi level. For simplicity, we take the same bandwidth and DOS for all the partial-wave components $\hat{l}$ of the conduction electrons. The RG equations, (\ref{RG1}) and (\ref{RG2}), are quite general and are not dependent on details of models.

In Table 1, IR, $\hat{\Psi}^{\gamma}$, describing hopping between the different  ion configurations for the six-level model, are listed. In six-level model, there are three IR's: $\G_1^+(\to a_1),\ \G_3^+(\to \{a_{3\up},a_{3\dw}\}=\{a_{3\ (3z^2-r^2)/\sqrt{3}},a_{3\ (x^2-y^2)}\})$ and $\G_4^-(\to \{a_{4x},\ a_{4y},\ a_{4z}\})$. The energy of the IR's are estimated as $\Delta_1<\Delta_4<\Delta_3$ for the case of retaining only the nearest neighbor hopping. For the conduction electrons, we retain the partial waves of $l\le 2$ and define $s\equiv\sum_k c_{k00},p_z\equiv\sum_k c_{k10},\ \sqrt{2}p_{x}\equiv\sum_k (c_{k1-1}- c_{k11}),\ \sqrt{2}p_{y}\equiv\sum_k {\rm i}(c_{k1-1}+ c_{k11}),\ \sqrt{2}d_{xy}\equiv\sum_k {\rm i}(c_{k2-2}-c_{k22}),\ \sqrt{2}d_{yz}\equiv\sum_k {\rm i}(c_{k21}+c_{k2-1}),\ \sqrt{2}d_{zx}\equiv\sum_k (c_{k2-1}-c_{k21}),\ e_{\up}\equiv\sum_k c_{k20}$, and $\sqrt{2}e_{\dw}\equiv\sum_k (c_{k22}+c_{k2-2})$. The IR's for the conduction electrons $\gamma_{msr}$ are listed in Table 2. Using $\hat{\bf \Psi}^{\gamma}$, $\gamma_{msr}$, we obtain the bare interaction Hamiltonian as

\begin{table}[t]

\caption{Definition of $\hat{\bf \Psi}^{\gamma}\equiv a_{\mu}^{\dg}{\bf \Psi}^{\gamma}_{\mu\nu}a_{\nu}$ for $(1,0,0)$ type cubic configurations. $n_{\mu} \equiv a^{\dg}_{\mu}a_{\mu}$, and $\hat{\bf \Psi}_{msr}^{(p)}$ is a $r$ component of $p$-th $\Gamma_{m}^{s}$ irreducible representation.}
	\begin{tabular}{rl}
         \hline
         \hline
 $\hat{\bf \Psi}_{1+}^{(1)}$ & $n_{1}$\\
 $\hat{\bf \Psi}_{1+}^{(2)}$ & $\frac{1}{\sqrt{2}}[n_{3\up}+n_{3\dw}]$\\
 $\hat{\bf \Psi}_{1+}^{(3)}$ & $\frac{1}{\sqrt{3}}[n_{4x}+n_{4y}+n_{4z}]$\\
 $\hat{\bf \Psi}_{2+}$       & $\frac{1}{\sqrt{2}}[a^{\dg}_{3\up}
a_{3\dw}-a^{\dg}_{3\dw}a_{3\up}]$ \\
 $\hat{\bf \Psi}_{3+\sigma}^{(1)}$ & $ \{ a^{\dg}_{1}a_{3\up},\ a^{\dg}_{1}a_{3\dw} \} $\\
 $\hat{\bf \Psi}_{3+\sigma}^{(2)}$ & $ \{ \frac{1}{\sqrt{2}}[n_{3\up}-n_{3\dw}], \ 
             -\frac{1}{\sqrt{2}}[a^{\dg}_{3\up}a_{3\dw}+a^{\dg}_{3\dw}a_{3\up}] \}$\\
 $\hat{\bf \Psi}_{3+\sigma}^{(3)}$ & $ \{ \frac{1}{\sqrt{6}}[n_{4x}+n_{4y}-2n_{4z}], \ 
             \frac{1}{\sqrt{2}}[-n_{4x}+n_{4y}]     \}$\\
 $\hat{\bf \Psi}_{4+\mu}$ & $ \{ \frac{1}{\sqrt{2}}[a^{\dg}_{4x}a_{4y}-{\rm h.c.}],
              \frac{1}{\sqrt{2}}[a^{\dg}_{4y}a_{4z}-{\rm h.c.}],
              \frac{1}{\sqrt{2}}[a^{\dg}_{4z}a_{4x}-{\rm h.c.}] \}$\\
 $\hat{\bf \Psi}_{5+\mu}$ & $ \{ \frac{1}{\sqrt{2}}[a^{\dg}_{4x}a_{4y}+{\rm h.c.}],
              \frac{1}{\sqrt{2}}[a^{\dg}_{4y}a_{4z}+{\rm h.c.}],
              \frac{1}{\sqrt{2}}[a^{\dg}_{4z}a_{4x}+{\rm h.c.}] \} $\\
 $ \hat{\bf \Psi}_{4-\mu}^{(1)}$ & $ \{ a^{\dg}_1a_{4x},\  a^{\dg}_{1}a_{4y},\  a^{\dg}_{1}a_{4z}\} $\\
 $ \hat{\bf \Psi}_{4-\mu}^{(2)}$ & $ \{ a^{\dg}_{4x}[\frac{\sqrt{3}}{2}a_{3\dw}-\frac{1}{2}a_{3\up}], \ 
              -a^{\dg}_{4y}[\frac{\sqrt{3}}{2}a_{3\dw}+\frac{1}{2}a_{3\up}],\ a^{\dg}_{4z}a_{3\up}\}$\\
 $\hat{{\bf \Psi}}_{5-\mu}$ & $ \{ a^{\dg}_{4z}a_{3\dw}, \ 
              \frac{1}{2}a^{\dg}_{4x}[\sqrt{3}a_{3\up}+a_{3\dw}],\
              \frac{1}{2}a^{\dg}_{4y}[\sqrt{3}a_{3\up}-a_{3\dw}] \}$\\
       \hline
         \hline
\end{tabular}

\label{tbl1}
\end{table}

\begin{table}[b!]

\caption{Definition of ${\gamma}^{(p)}_{msr}$, $r$ component of $p$-th irreducible representation $\Gamma_{m}^{s}$.}
	\begin{tabular}{rl}
         \hline
         \hline
$\gamma_{1+}^{(1)}$ & $s^{\dg}s$\\
$\gamma_{1+}^{(2)}$ & $p_x^{\dg}p_x+p_y^{\dg}p_y+p_z^{\dg}p_z$\\
$\gamma_{1+}^{(3)}$ & $e_{\up}^{\dg}e_{\up}+e_{\dw}^{\dg}e_{\dw}$\\
$\gamma_{1+}^{(4)}$ & $d_{xy}^{\dg}d_{xy}+d_{yz}^{\dg}d_{yz}+d_{zx}^{\dg}d_{zx}$\\
$\gamma_{2+}$ & $\frac{1}{2}[e_{\up}^{\dg}e_{\dw}-e_{\dw}^{\dg}e_{\up}]$\\
$\gamma_{3+\sigma}^{(1)}$ & $ \{ s^{\dg}e_{\up},\ s^{\dg}e_{\dw} \}$\\
$\gamma_{3+\sigma}^{(2)}$ & $ \{ \frac{1}{2}[e_{\up}^{\dg}e_{\up}-e_{\dw}^{\dg}e_{\dw}], \ -\frac{1}{2}[e_{\up}^{\dg}e_{\dw}+e_{\dw}^{\dg}e_{\up}]     \}$ \\
$\gamma_{3+\sigma}^{(3)}$ & $ \{ \frac{1}{\sqrt{12}}[p_{x}^{\dg}p_{x}+p^{\dg}_{y}p_{y}-2p_z^{\dg}p_z], \ \frac{1}{2}[-p_{x}^{\dg}p_{x}+p_{y}^{\dg}p_{y}]     \}$ \\
$\gamma_{3+\sigma}^{(4)}$ & $ \{ \frac{1}{\sqrt{12}}[2d_{xy}^{\dg}d_{xy}-d_{yz}^{\dg}d_{yz}-d_{zx}^{\dg}d_{zx}],\  \frac{1}{2}[d_{yz}^{\dg}d_{yz}-d^{\dg}_{zx}d_{zx}]  \}$ \\
$\gamma_{4+\mu}^{(1)}$ & $ \{ \frac{1}{2}[p_{x}^{\dg}p_{y}-{\rm h.c.}], \ 
              \frac{1}{2}[p_{y}^{\dg}p_{z}-{\rm h.c.}],\
              \frac{1}{2}[p_{z}^{\dg}p_{x}-{\rm h.c.}]          \}$\\
$\gamma_{4+\mu}^{(2)}$ & $ \{ \frac{1}{2}[d^{\dg}_{yz}d_{zx}-{\rm h.c.}], \ 
              \frac{1}{2}[d_{zx}^{\dg}d_{xy}-{\rm h.c.}],\
              \frac{1}{2}[d_{xy}^{\dg}d_{yz}-{\rm h.c.}]          \}$\\
$\gamma_{4+\mu}^{(3)}$ & $ \{ e^{\dg}_{\dw}d_{xy}, \ 
              -\frac{1}{2}[e_{\up}^{\dg}+\sqrt{3}e^{\dg}_{\dw}]d_{yz},        \     \frac{1}{2}[\sqrt{3}e_{\dw}^{\dg}-e_{\up}^{\dg}]d_{zx} \}$\\
$\gamma_{5+\mu}^{(1)}$ & $ \{ s^{\dg}d_{xy},\  s^{\dg}d_{yz},\  s^{\dg}d_{zx}\}$\\
$\gamma_{5+\mu}^{(2)}$ & $ \{ \frac{1}{2}[p_{x}^{\dg}p_{y}+{\rm h.c.}], \ 
              \frac{1}{2}[p_{y}^{\dg}p_{z}+{\rm h.c.}],\
              \frac{1}{2}[p_{z}^{\dg}p_{x}+{\rm h.c.}]          \}$\\
$\gamma_{5+\mu}^{(3)}$ & $  \{ \frac{1}{2}[d^{\dg}_{yz}d_{zx}+{\rm h.c.}], \ 
              \frac{1}{2}[d_{zx}^{\dg}d_{xy}+{\rm h.c.}],\
              \frac{1}{2}[d_{xy}^{\dg}d_{yz}+{\rm h.c.}]          \}$\\
$\gamma_{5+\mu}^{(4)}$ & $ \{ e^{\dg}_{\up}d_{xy}, \ 
           \frac{1}{2}[\sqrt{3}e_{\up}^{\dg}-e_{\dw}^{\dg}]d_{yz},\ 
            -\frac{1}{2}[\sqrt{3}e_{\up}^{\dg}+e_{\dw}^{\dg}]d_{zx}
               \}$\\
$\gamma_{2-}$ & $ [p_{x}^{\dg}d_{yz}+p_{y}^{\dg}d_{zx}+p_{z}^{\dg}d_{xy}]$\\
$\gamma_{3-\sigma}$ & $ \{ \frac{1}{2}[p_{x}^{\dg}d_{yz}-p_{y}^{\dg}d_{zx}],\ \frac{1}{2}[p_{y}^{\dg}d_{zx}-p_{z}^{\dg}d_{xy}]\}$\\
$\gamma_{4-\mu}^{(1)}$ & $ \{ s^{\dg}p_{x},\  s^{\dg}p_{y},\  s^{\dg}p_{z}\}$
\\
$\gamma_{4-\mu}^{(2)}$ & $ \{ \frac{1}{2}[p_{x}^{\dg}[\sqrt{3}e_{\dw}-e_{\up}],\ -p_{y}^{\dg}[\sqrt{3}e_{\dw}+e_{\up}],p_{z}^{\dg}e_{\up}\}$\\
$\gamma_{4-\mu}^{(3)}$ & $ \{ \frac{1}{2}[p_{y}^{\dg}d_{xy}+p_{z}^{\dg}d_{zx}],
              \frac{1}{2}[p_{z}^{\dg}d_{yz}+p_{x}^{\dg}d_{xy}],
              \frac{1}{2}[p^{\dg}_{x}d_{zx}+p_{y}^{\dg}d_{yz}] \}$\\
$\gamma_{5-\mu}^{(1)}$ & $ \{ p_{z}^{\dg}e_{\dw}, \ 
              \frac{1}{2}p_{x}^{\dg}[e_{\dw}+\sqrt{3}e_{\up}],\
              \frac{1}{2}p_{y}^{\dg}[\sqrt{3}e_{\up}-e_{\dw}] \}$\\
$\gamma_{5-\mu}^{(2)}$ & $\{
             \frac{1}{2}[p^{\dg}_{x}d_{zx}-p_{y}^{\dg}d_{yz}],             
             \frac{1}{2}[p_{x}^{\dg}d_{xy}-p_{z}^{\dg}d_{yz}],
             \frac{1}{2}[p_{y}^{\dg}d_{xy}-p_{z}^{\dg}d_{zx}]            
              \}$\\
         \hline
         \hline
\end{tabular}

\label{tbl2}
\end{table}

\begin{eqnarray}
 \frac{H_{\rm int}}{u_0} \!\!\!\!\!&=& \!\!\!\!\!\frac{a^2}{3\sqrt{10}}\Big\{(\gamma_{3+\sigma}^{(1)}+{\rm h.c.}) +2\gamma_{3+\sigma}^{(3)}\Big \}[{\bf \Psi}_{3+\sigma}^{(1)}+{\rm h.c.}]\nonumber\\
&+& \!\!\!\!\!a^2\Big \{ \frac{1}{3\sqrt{10}}(\gamma_{3+\sigma}^{(1)}+{\rm h.c.}) +\frac{2}{3\sqrt{6}}\gamma_{3+\sigma}^{(3)}\Big\}{\bf \Psi}_{3+\sigma}^{(2)}\nonumber\\
&-& \!\!\!\!\!\sqrt{3}a^2\Big\{\frac{1}{3\sqrt{10}}(\gamma_{3+\sigma}^{(1)}+{\rm h.c.})+\frac{2}{3\sqrt{6}}\gamma_{3+\sigma}^{(3)}\Big\}{\bf \Psi}_{3+\sigma}^{(3)}\nonumber\\
&-& \!\!\!\!\!\frac{a}{\sqrt{3}}(\gamma_{4-\mu}^{(1)}+{\rm h.c.})[{\bf \Psi}_{4-\mu}^{(1)}+\sqrt{\frac{2}{3}}{\bf \Psi}_{4-\mu}^{(2)}+{\rm h.c.}] \nonumber\\
 &+& \!\!\!\!\!{a^2e^{-\frac{a^2}{2\sigma^2}}}\Bigg[\Big(\frac{2}{3}\gamma_{1+}^{(1)}+\frac{2}{9}\gamma_{1+}^{(2)}\Big){\bf \Psi}_{1+}^{(1)}\nonumber\\
&&-\sqrt{2}\Big(\frac{1}{3}\gamma_{1+}^{(1)}+\frac{1}{9}\gamma_{1+}^{(2)}\Big){\bf \Psi}_{1+}^{(2)}\nonumber\\
&-&\!\!\!\!\!\Big\{\frac{1}{6\sqrt{10}}(\gamma_{3+\sigma}^{(1)}+{\rm h.c.})+\frac{1}{3\sqrt{6}}\gamma_{3+\sigma}^{(3)}\Big\}[{\bf \Psi}_{3+\sigma}^{(1)}+{\rm h.c.}]\nonumber\\
&+&\!\!\!\!\!\Big\{\frac{1}{3\sqrt{10}}(\gamma_{3+\sigma}^{(1)}+{\rm h.c.})+\frac{2}{3\sqrt{6}}\gamma_{3+\sigma}^{(3)}\Big\}[{\bf \Psi}_{3+\sigma}^{(2)}+{\rm h.c.}]\nonumber\\
&+& \!\!\!\!\!\Big\{\frac{1}{\sqrt{30}}(\gamma_{5+\mu}^{(1)}+{\rm h.c.})+\frac{\sqrt{2}}{3}\gamma_{5+\mu}^{(2)}\Big\}{\bf \Psi}_{5+\mu}\Bigg],\label{Hint}
\end{eqnarray}
where we have used approximations $a\ll 1$ and a local interaction parameterized by $u_0$\cite{HHM}. The initial couplings ${J_{ij}^{\gamma}}^{(0)}$'s in the sense of RG are estimated from eq. (\ref{Hint}). 

\begin{figure}[t]
  \begin{center}
    \includegraphics[width=.47\textwidth]{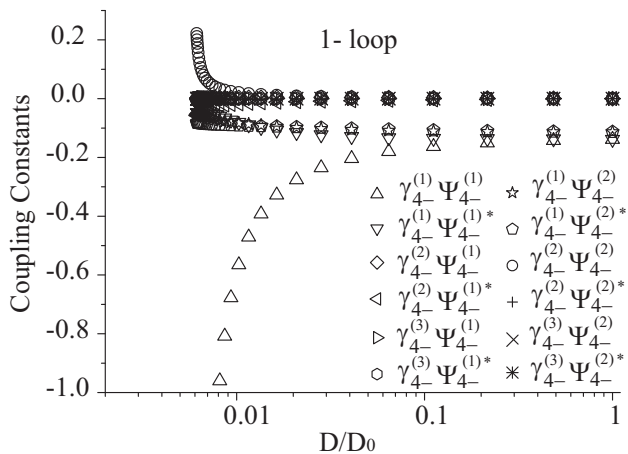}
  \end{center}
\caption{1-loop RG flows. The parameters used are $u_0=0.3D_0$, $a=0.8k_F^{-1}$ and $\sigma=0.28k_F^{-1}$.}
\label{fig-1}
\end{figure}

\begin{figure}[t]
  \begin{center}
    \includegraphics[width=.47\textwidth]{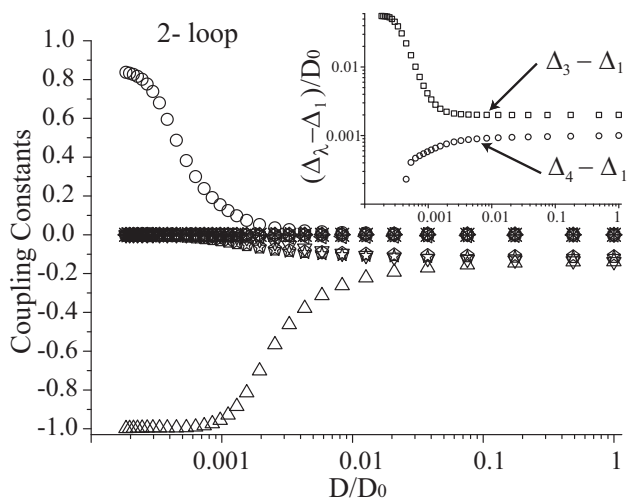}
  \end{center}
\caption{2-loop RG flows. The symbols and parameters are same as in Fig. {\ref{fig-1}}. Inset: Effective levels of the ion states $\Delta_{\mu}/D_0$ vs $D/D_0$, $\mu=1,3$ and $4$.}
\label{fig-2}
\end{figure}

We solve the RG equations (\ref{RG1}) and (\ref{RG2}) numerically.
The results of 1-loop and 2-loop RG are shown in Fig. \ref{fig-1}  and Fig. \ref{fig-2}, respectively (we only show for typical couplings). The qualitative features are similar to the case of the four-level model\cite{HHM}. In 1-loop calculation, we obtain the characteristic energy scale $T_K/D_0\simeq \exp(-1/(\rho u_0))$, at which some of the effective coupling constants diverge. In the 2-loop order, the coupling constants saturate at intermediate values approaching the fixed point. What we have to be careful in Fig. {\ref{fig-2}} is that eqs. (\ref{RG1}) and (\ref{RG2}) are derived assuming $D\gg \max(J_{ij}^{\gamma},\ \Delta_{\mu})$. Then the reliable range of $D/D_0$ is, at least, $D/D_0 > \Delta/D_0\simeq 0.001$. It is noted that the energy of the first excited level of the ion decreases as renormalization proceeds, while that of the second one increases.

\begin{figure}[t!]
  \begin{center}
    \includegraphics[width=.48\textwidth]{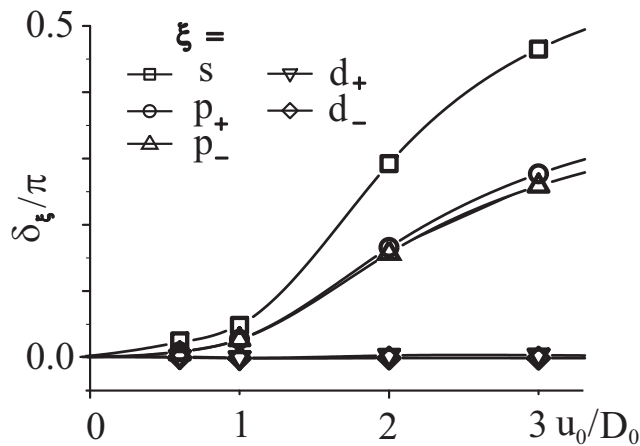}
  \end{center}
\caption{Phase shift of conduction electrons $\delta_{\xi}$ vs $u_0$. $\xi$ indicates the orbital index of the conduction electron: $p_+\equiv-(p_x+{\rm i}p_y)/\sqrt{2}$, $p_-\equiv(p_x-{\rm i}p_y)/\sqrt{2}$, $d_+\equiv (e_{\dw}+{\rm i}d_{xy})/\sqrt{2}$ and $d_-\equiv(e_{\dw}-{\rm i}d_{xy})/\sqrt{2}$. Lines are guide to eyes.}
\label{fig-3}
\end{figure}

In the case of four-level system, the correct fixed point corresponds to the results obtained by 1-loop calculation, which was verified by the NRG calculation. The detailed analysis shows that the $s,\ p_x\ {\rm and}\ p_y$ components of the conduction electrons form Kondo singlet with the ion with $\Gamma_1^+$ (singlet) and $\Gamma_5^{-}$ (doublet) symmetry under $D_{4h}$ point group. This can be seen in Fig. \ref{fig-3}, in which we show the phase shifts of the conduction electrons estimated by the NRG energy spectra. The orbitals of the $s$ and $p_{\pm}$ strongly interact with the ion, while those of $d_{\pm}$ do not. These kinds of orbital Kondo effect seem to give an origin of the heavy effective mass observed in SmOs$_4$Sb$_{12}$. Indeed, we obtained the enhanced linear temperature coefficient of the specific heat in this model by the NRG calculations.\cite{HHM}

 In the case of six-level system, the dominant coupling constants at low energy fixed point are those between $(s,\ p_x\ p_y,\ p_z)$, and $(\Gamma_1^+,\ \Gamma_{4x}^-,\ \Gamma_{4y}^-,\ \Gamma_{4z}^-)$, which is a natural extension of the results in the four-level system. At low temperatures, an orbital Kondo effect arising from these couplings are thought to become possible. 
Although we cannot carry out the NRG calculation for the six-level case because of the computational difficulty, we expect that the result of the 1-loop calculation captures the nature of the low energy fixed point as the four-level model. 

In the present paper, we have investigated $(1,0,0)$ type six-level model. Recently, Kaneko et al. carried out neutron scattering experiments in PrOs$_4$Sb$_{12}$, and deduced the nuclear density distributions.\cite{Kaneko} Their result suggests that the charge distribution of Pr ion extends mainly in the $(1,1,1)$ direction at high temperatures and isotropically at low temperatures. Concerning this point, it is desired to investigate the properties of an eight-level model, in which the stable points of the ion locate at eight (1,1,1) directions. We expect that the similar orbital Kondo effects occur even in the eight-level model.


\vspace{1cm}
We would like to thank Y. Nemoto, Y. Nakanishi, K. Kaneko and T. Goto for fruitful discussions. One of us (K. H.) is supported by Research Fellowships of JSPS for Young Scientists.
This work is supported by a Grant-in-Aid for Scientific Research (No.16340103), 21st Century COE Program (G18) by Japan Society for the Promotion of Science, and a Grant-in-Aid for Scientific Research in Priority Areas (No. 16037209) by MEXT.

\end{document}